\documentclass[a4paper,11pt]{article}

\usepackage{paper}

\usepackage[T1]{fontenc} 
\usepackage{amsmath}
\usepackage{graphicx}
\usepackage{booktabs}
\usepackage{multirow}
\usepackage{xspace}
\usepackage{subfigure,graphicx}

\title{\boldmath Remarks on the numerical impact of potential
  theoretical systematics in the prediction of QCD instanton cross
  sections} 

\author[a]{Michelangelo~L.~Mangano}

\affiliation[b]{CERN TH Department, CH-1211 Geneva 23, Switzerland}

\emailAdd{michelangelo.mangano@cern.ch}

\abstract{{We discuss the origin and size of potential uncertainties
    arising in the estimate of cross sections for the production of
    multiparticle final states induced by QCD instantons at the LHC.}  }

\begin{document}
\maketitle

\def\be{\begin{equation}}
\def\ee{\end{equation}}
\def\K{Khoze:2020tpp}
\def\KK{Khoze:2019jta}
\def\M{Mueller:1990qa}
\def\MM{Mueller:1990ed}
\def\MMM{Mueller:1991fa}
\def\amp{ {\cal{A}} }
\def\rhobar{ \bar{\rho} }
\def\asrho{ \alpha_s(1/\rho) }
\def\asrhobar{ \alpha_s(1/\rhobar) }
\def\as{ \alpha_s }
\def\bvskip{ \vskip 5mm
 \hrule
 \vskip 2mm
}
\def\evskip{ \vskip 2mm
 \hrule
 \vskip 5mm
}
\section{Introduction}
This note is a more detailed version of a contribtion to the ``QCD Instantons'' discussion session at
the Workshop on ``Topological Effects in the Standard Model:
Instantons, Sphalerons and Beyond at LHC'',
\url{https://indico.cern.ch/event/965112/}. It follows the two
presentations by A.Ringwald and V.Khoze, and in particular it links
to the latter talk, and to the recent papers on which this
was based, refs.~\cite{\K,\KK}. It incorporates some additional
remarks emerged during the discussions at the Workshop, and further follow up.

This note will not address the underlying formalism used and developed
in refs.~\cite{\K,\KK}. Taking the formalism and the results for
granted, I focus on possible systematics that should still be attached
to the results in ~\cite{\K,\KK}, helping to put in perspective the
interpretation of the LHC searches for final states induced by QCD
instantons.

\section{Results and remarks}
The amplitude for the instanton-induced production of $n_g$ gluons and
$n_f$ quark-antiquark pairs is given in ref.~\cite{\K} as:
\be \label{eq:amp}
  \amp(2\to n_g+2n_f) \sim \int_0^\infty \, d\rho^2 \left( \rho^2
  \right)^{n_g+n_f-1} \; e^{-\frac{2\pi}{\asrho} -
    \frac{\asrho}{16\pi}E^2 \rho^2 \log(E^2\rho^2) }
  \ee
where $\rho$ is the instanton radius, and $E=\sqrt{\hat{s}}$ is the
partonic CM energy.  As in ref.~\cite{\K}, we neglect overall constants, wave function
normalization factors, etc. The second term in the exponent,
proportional to $\rho^2\log(E^2\rho^2)$, reflects the Mueller's
form-factor, evaluated and discussed in refs.~\cite{\M,\MM,\MMM}. This
is critical to suppress the contribution of large-size instantons, and
to guarantee the convergence of the integral over instanton
configurations. We note that eq.~\ref{eq:amp} makes explicit use of
the relation $\rho\,\mu_r=1$ between the renormalization scale $\mu_r$
(which otherwise enters in the value of the strong coupling constant,
in the instanton density
and in Mueller's exponent) and the radius $\rho$~\cite{\K}. This relation
enforces the renormalization group invariance of the amplitude, and
reflects the intuitive notion that the inverse size of the instanton
defines the proper scale at which strong interactions act in the
instanton field; as
discussed in the following, however, the relation should be taken only
as a functional relation fulfilling scale invariance, leaving room for
a possible overall numerical factor, namely  $\rho \, \mu_r={\cal{O}}(1)$.

\bvskip

\noindent
    {\bf Remark 1}.   From dimensional analysis, the amplitude given above scales with $E$
  as follows:
  \be
  \amp \propto E^{-2(n_g+n_f+b_0/2)} \; \propto \; E^{-2(n_g+n_f)} \;
  \left( \frac{\Lambda}{E}\right)^{b_0} \; ,
  \ee
  where at leading order $\alpha_s(\mu)=4\pi/[b_o\,\log(\mu^2/\Lambda^2)]$ and $b_0=11-2/3\, n_f$.
  The second expression above highlights the fact that, while $(n_f+n_g)$
  powers of $1/E^2$ are matched by the energy dependence of
  the wave function normalization of the external states and by the
  final phase-space integration, leading to dim($\sigma)=-2$, the
  $b_0$ powers of $1/E$ are matched instead by
  the QCD scale $\Lambda$, which therefore must appear in the final
  amplitude expression. This is the consequence of the power
  suppressed nature of this non-perturbative amplitude, embodied by
  the contribution of the instanton action $\exp(-2\pi/\asrho)$. This means that the amplitude
  has an intrinsic $\Lambda^{b_0}$ dependence. More on this later.  

  \evskip
  
\noindent As indicated in ref.~\cite{\K}, the amplitude can be evaluated in the
 saddle-point approximation, where, leaving out again constant numerical factors:
 \be \label{eq:intf}
 \amp \sim \int_0^\infty \, d\rho^2 e^{f(\rho^2)} \; = \;
 e^{f(\rhobar)}\sqrt{ \frac{2\pi}{-f''(\rhobar)}  }
 \ee
 where 
 \be
 f(\rho^2)=(n_g+n_f-1+b_0/2) \, \log\rho^2 \, - \,
 \frac{\asrho}{16\pi}E^2\rho^2\log(E^2\rho^2) \; ,
 \ee
 The saddle point $\rhobar$ is defined through:
 \be \label{eq:saddle}
 \partial f(\rho^2)/\partial \rho^2 = \frac{A}{\rho^2} -\frac{E^2}{16\pi}
 \asrho \log(E^2\rho^2) + {\cal{O}}(\alpha_s^2) =0 \quad \mathrm{at} \quad \rho=\rhobar
 \ee
 where $A=n_g+n_f-1+b_0/2$. We used the a-posteriori knowledge that
 $E^2\rhobar^2 \gg 1$, to neglect in the derivative a term of order 1 w.r.t. $\log(E^2\rho^2$).

 \bvskip

\noindent
 {\bf Remark 2}. Notice that in eq.~\ref{eq:saddle} we neglected terms
 formally of higher powers of $\alpha_s$. Some arise by including the
 term proportional to $\partial \asrho/\partial\rho^2 \,\sim \, b_0 \as^2(1/\rho)$ in the
 derivative of $f(\rho)$, others would arise in taking the NLO
 beta-function rather than the LO one; others, unknown, arise from NLO
 corrections to the function $f(\rho)$ itself, in particular to
 Mueller's form factor. Since we cannot control
 the exact form of these higher order terms, we stick to the strict LO
 expression, but must keep in mind their existence for the assessment
 of the systematics of the final result.
 \evskip
 
 \noindent
The approximate solution of the saddle-point condition leads to the relation:
 \be \label{eq:rhobar}
 \frac{1}{\rhobar^2} \; = \; \eta \, E^2 \, \as(\eta E^2) \; \log(1/\eta) 
   + {\cal{O}}(\as^2) 
 \ee
 where $\eta=1/(16\pi A)$. This approximate solution to
 eq.~\ref{eq:saddle} agrees to within 10\% with the exact solution,
 estimated numerically. Equation~\ref{eq:rhobar} 
 leads to a value of the inverse instanton
 radius of $\epsilon = 1/\rhobar = \gamma E$, with $\gamma$ in the range of
 $1/20-1/30$ for $E\sim 100-3000$~GeV. This is consistent with the
 findings of Fig.~4 and Table~1 of ref.~\cite{\KK}. The fact that the
 inverse instanton radius is significantly larger than the invariant
 mass of the system, $E$, has important implications. On one side it
 sets a threshold for the creation of massive quarks Q in the instanton
 decay: for the instanton to resolve the heavy quark, it must be
 $\rho<1/m_Q$, and therefore $E>m_Q/\eta \sim 30 m_Q$. This means that
 the threshold to produce bottom quarks is at around 150~GeV, and to
 produce top quarks it must be $E\gtrsim 5$~TeV. On the other side, it
 implies that for $\alpha_s$ to remain in the perturbative domain,
 $1/\rho > \Lambda$, the minimum energy should be $E>\Lambda/\eta\sim
 4$~GeV.

 Evaluating $f''(\rho)$ at the saddle point gives
 \be
 f''(\rhobar)=-\frac{A}{\rhobar^4} ( 1 +  {\cal{O}}(\as) )
 \ee
 and, putting things back into eq.~\ref{eq:intf} and neglecting overall
 constant factors, we obtain:
 \be \label{eq:final}
 \amp \sim \left(\frac{\Lambda}{E}\right)^{b_0} \;
 \left(\frac{1}{E}\right)^{n_g+n_f} \; \left [ \frac{1}{\as(\eta E^2)
     [1+O(\as)]} \right]^{n_g+n_f+b_0/2} \; .
 \ee

\section{Discussion}
I discuss here in more detail the numerical impact on the final
amplitude, eq.~\ref{eq:final}, of the possible sources of systematics
highlighted so far.

\subsection{$\Lambda^{b_0}$}
The first point is the $\Lambda^{b_0}$ term upfront. On one side this
inherits the intrinsic 1\% uncertainty on $\as(M_Z)$. But
$\Delta\Lambda/\Lambda\sim \Delta\as/\as \times \log(M_Z/\Lambda) \sim
6\%$, leading to $\Delta\amp/\amp \sim \pm 60\%$, which is negligible
overall. On the other hand, the choice of the perturbative order at
which $\Lambda$ is estimated is not well defined, and e.g. the difference
between $\Lambda_{LO}$ and $\Lambda_{NLO}$ is large. For example, to
obtain $\as(M_Z)=0.12$ from the 1-loop evolution we get $\Lambda\sim
100$~MeV, while at 2-loop we get $\Lambda\sim 260$~MeV. So, in the
cross section $\sigma\propto \amp^2$ there is a potential systematics
factor in the range of $2^{\pm b_0} \sim [4\times 10^{-3} - 250]$.

\subsection{NLO effects}
There is an independent uncertainty arising from the $O(\as)$
corrections indicated in eq.~\ref{eq:final}. This is independent of
the LO vs NLO issue raised in the previous remark: there we dealt with
the order at which the leading power-suppressed instanton action,
$\exp(-2\pi/\as)$, is calculated. Here we are dealing with
higher-order corrections to Mueller's form factor. It is reasonable to
expect these $O(\as)$ uncertainties to be limited to a $\pm 20\%$, but
when raised to the power of $(n_g+n_f+b_0/2)$ this can become an
overall factor of $(1.2/0.8)^{(n_g+n_f+b_0/2)} \geq 50$ for the
amplitude, and greatly more for the cross sections.  More in general,
as mentioned above, the relation between renormalization scale $\mu_r$ and the instanton
radius, $\rho \, \mu_r  =1$, should be subject to the usual
factor of 2 uncertanity. The legitimate choice of $\mu_r \rho
=[0.5-2]$ would lead to a factor of $[0.5-2]$ rescaling of $\Lambda$
in the argument of $\as$, leading again to a systematics similar to
what discussed at point 1 above.

\begin{table}
\centering
\caption{Parton luminosity ratios evaluated at $\mu_F=E$ and
  $\mu_F=1/\rho$, at different partonic CM energies $E$. For each $E$,
  the corresponding value of $1/\rho$ is derived by linear
  interpolation from Table~1 of
  ref.~\cite{\KK}, subject to the additional constraint $\mu_F \ge 1.65$~GeV.}
\vskip 2mm
\begin{tabular}{l|ccccccc}
  $E$ (GeV)  & 20 & 30 & 40 & 50 & 100 & 200 & 500 \\
  $1/\rho$ (GeV)  & 1.65 & 2.1 & 2.7 & 3.2 & 5.4 & 9.8 & 22 \\
  \hline
  $\left[ d{\cal{L}}/d\tau \right]_{(\mu_F=E)} \; / \; \left [
    d{\cal{L}}/d\tau \right]_{(\mu_F=1/\rho)}$
  & 49 & 15 & 7.4 & 5.0 & 2.1 & 1.2 & 0.8 \\
\end{tabular}
\label{tab:lumi}
\end{table}

\begin{table}
\centering
\caption{Cross sections, at 13 TeV, for the
  production of QCD instantons with mass larger than $E_{min}$. First row:
  the results from Table~2 of ref.~\cite{\KK}. (a) My result, mimiking the prescriptions in
  ref.~\cite{\KK}, including $\mu_F=1/\rho$. (b) Same approach, but
  with $\mu_F=E$. In rows (c) and (d) I repeat the calculation
  of (a) and (b), 
  using a local power-like interpolation for the partonic
  cross section, instead of the linear
  interpolation adopted in~\cite{\KK}.  Notice that, while significantly reduced with respect to the
  result of the linear interpolation, the rate for
  $E>20$~GeV and with $\mu_F=E$ is still larger than the total pp cross section. }
\vskip 2mm
\begin{tabular}{l|ccccccc}
$E_{min}$ (GeV)  & 20 & 30 & 40 & 50 & 100 & 200 & 500 \\ \hline
$\sigma(pp\to I)$~\cite{\KK} & 6.3\,mb & & & 41\,$\mu$b & 80\,nb & 105\,pb
  & 3.5\,nb \\

$\sigma(pp\to I)$~(a) & 5.8\,mb & 0.91\,mb & 0.17\,mb & 40\,$\mu$b &
  79\,nb & 106\,pb & 3.5\,fb \\

$\sigma(pp\to I)$~(b) & 170\,mb & 9.1\,mb & 0.9\,mb & 0.2\,mb &
  150\,nb & 120\,pb & 2.5\,fb \\
 
$\sigma(pp\to I)$~(c) & 4.0\,mb & 0.63\,mb & 0.12\,mb & 25\,$\mu$b &
  41\,nb & 39\,pb & 2.0\,fb \\

  $\sigma(pp\to I)$~(d) & 110\,mb & 6.5\,mb & 0.71\,mb & 0.11\,mb &
  80\,nb & 46\,pb & 1.4\,fb \\
\end{tabular}
\label{tab:xs}
\end{table}

  \subsection{Factorization scale systematics}
When the partonic cross sections are convoluted with parton
luminosities, to extract hadronic cross sections in pp collisions, a
further source of systematics arises from the choice of the
factorization scale, $\mu_F$. In general, this is set equal to the
renormalization scale, which for the instanton case, as discussed
above, is chosen around the value of $1/\rho \ll E$. Since the
factorization scale is related to the removal of initial state
collinear singularities, which factorize out of the details of the
hard process itself, it is fair to argue however that the choice
$\mu_F \sim E$, if not preferable, should at least be considered here
as a possible alternative. One can anticipate that this could lead to
a large uncertainty, since the partonic luminosity at the small values
of $\mu_F$ probed by the choice $\mu_F=1/\rho$ has a very strong scale
dependence, as shown in Table~\ref{tab:lumi}.
   
Table~\ref{tab:xs} shows the actual impact of this systematics from
the choice of $\mu_F$. The first row of the table shows the results of
ref.~\cite{\KK}: these were obtained by convoluting the parton level
cross section $\hat\sigma(E)$ with
the partonic luminosity at $\mu_F= {\mathrm{max}} [1/\rho,Q_{min}]$, $Q_{min}=1.65$~GeV being the
minimum value of $Q$ supported by the NNPDF3.1 PDF set chosen for the
calculation.  For simplicity, ref.~\cite{\KK} obtained $\hat\sigma(E)$
from a linear interpolation of the values calculated for a few fixed
choices of the instanton mass $E$, given in Table~1 of
ref.~\cite{\KK}~\footnote{I remark that, given the rapid falloff of
  the cross section, which for $E\gtrsim 50$~GeV behaves like
  $1/E^{\sim 9}$, a bin-by-bin linear interpolation of
  $\log\hat\sigma(E)$, namely a power fit in $E$ of $\hat\sigma(E)$
  rather than a linear interpolation, would give a more faithful
  representation.}.

The second row in the Table is what I get by replicating the
calculation of ref.~\cite{\KK}. The results show good agreement, with
minor differences likely due to the different implementations of the
interpolation criterion (I assume also a linear interpolation for the
values of $\rho$ at different $E$ values in the integration). Having
estabished that I can reproduce the results of ref.~\cite{\KK}, the third
row shows the results I obtain by setting $\mu_F=E$, the instanton
mass. The larger evolution of the PDFs in $Q^2$ leads to a drastic
increase of the cross section at small $E$ values, driven by the
low-$x$ behaviour of the gluon density. At larger $E$, the $Q^2$
evolution depletes the gluon at larger $x$ values, leading to a
reduced rate. This implies that it is not just the absolute value of
the cross sections that is affected by potentially large
uncertainties, but also the shape of the instanton-energy dependence
seems to be uncertain, making it less useful in separating signal and
backgrounds or in the interpretation of potential signals.  For
example, in the region of interest for the extraction of a signal,
Table~\ref{tab:xs} shows that the drop in rate between
$\sigma(E_{min}=50$\,GeV) and $\sigma(E_{min}=500$\,GeV) is 7 times
larger with the choice $\mu_F=E$ than with $\mu_F=1/\rho$.

We also note that, for $E_{min}=20$~GeV, the total cross section
largely surpasses the total pp cross section, more than doubling the
inelastic rate. This apparent breaking of unitarity would simply imply
that the average number of QCD instantons with $E>20$~GeV produced in
a pp collision is larger than 1~\footnote{A similar phenomenon occurs when we
calculate the cross section for production of minijets with a low
minimum $p_T$. Notice also that for $\sqrt{ \hat{s}} \sim 20 $~GeV at the LHC
  one does not expect gluon-shadowing and saturation effects to play a
  unitarization role, since there is no evidence of gluon saturation
  from the measurement of other hard processes at this $(x,Q)$ scale
  (e.g. charm and bottom production at small $p_T$).}, but would also
imply that the total pp cross section is saturated by instanton
production, a rather unlikely possibility. Among other effects, this would lead
to the striking prediction of an average multiplicity of charm
quark pairs produced per pp collisions larger than 1 (the standard charm
production mechanism predicts a total NLO rate of about 10\,mb, although with
a large uncertainty. This corresponds to about one charm pair produced
for every 8 inelastic events).

 \section{Conclusions}
In conclusion, it appears that there could be large sources of
systematics associated to the predictions for instanton-induced QCD
processes at the LHC. If the analysis reported in this note is
correct, these uncertainties could cover several orders of
magnitude. This does not remove interest in the search for such final
states, but a possible lack of evidence does not lead to the immediate
conclusion that instantons ``do not exist'', but simply that their
actual production rate is unfortunately on the lower end of the
systematics, with respect to the central baseline rates discussed in
refs.~\cite{\K,\KK}. These uncertainties would also clearly influence
the interpretation of a possible signal, and its clear identification
in terms of instantons, as opposed to other possible sources, within
or beyond the Standard Model.

\bibliographystyle{paper}
\bibliography{paper}

\end{document}